# From Drone Photographs to Livability Mapping: AI-powered Environment Perception in Rural China

Weihuan Deng [a,b,c], Yaofu Huang [a,b,c], Luan Chen [a,b,c] , Xun Li [a,b,c*], Yu Gu [a,b,c], Yao Yao [d]

[a] *School of Geography and Planning, Sun Yat-Sen University, Guangzhou 510275, Guangdong Province, China*

*Corresponding author: lixun@mail.sysu.edu.cn

[b] *China Regional Coordinated Development and Rural Construction Institute, Urbanization Institute, Sun Yat-Sen University, Guangzhou 510275, Guangdong Province, China*

[c] *Key Laboratory of Rural Digital Technology, Ministry of Housing and Urban-Rural Development, Guangzhou 510275, Guangdong Province, China*

[d] *School of Geography and Information Engineering, China University of Geosciences, Wuhan 430078, Hubei Province, China*

## Abstract

The high cost of acquiring rural street view images has constrained comprehensive environmental perception in rural areas. Drone photographs, with their advantages of easy acquisition, broad coverage, and high spatial resolution, offer a viable approach for large-scale rural environmental perception. However, a systematic methodology for identifying key environmental elements from drone photographs and quantifying their impact on environmental perception remains lacking. To address this gap, a Vision-Language Contrastive Ranking Framework (VLCR) is designed for rural livability assessment in China. The framework employs chain-of-thought prompting strategies to guide multimodal large language models (MLLMs) in identifying visual features related to quality of life and ecological habitability from drone photographs. Subsequently, to address the instability in pairwise village comparison, a text description-constrained drone photograph comparison strategy is proposed. Finally, to overcome the efficiency bottleneck in nationwide pairwise village comparisons, an innovation ranking algorithm based on binary search interpolation is developed, which reduces the number of comparisons through automated selection of comparison targets. The proposed framework achieves superior performance with a Spearman Footrule distance of 0.74, outperforming mainstream commercial MLLMs by approximately 0.1. Moreover, the mechanism of concurrent comparison and ranking demonstrates a threefold enhancement in computational efficiency. Our framework has achieved data innovation and methodological breakthroughs in village livability assessment, providing strong support for large-scale village livability analysis.

## 1. Introduction

With the rapid development of mapping services and crowdsourcing platforms, visual data represented by street-view imagery has experienced explosive growth (Cheng et al. 2018; Danish et al. 2025). As precise spatial data covering the road network, street-view images not only authentically document the physical environment of a region (Huang et al. 2024), including building facades, road infrastructure, and green landscape, but also contain in-depth information about regional functional layouts and social activities (Gao et al. 2023). These extensive, timely, and information-rich visual data, combined with advances in deep learning technologies, are transforming traditional research paradigms (Huang et al. 2023). They provide new data support and analytical approaches for physical environmental perception, environmental quality assessment, human settlement environment monitoring, and other fields (Wei et al. 2022; Huang et al. 2025).

Environmental perception studies based on street-view imagery originated from Salesses et al. (2013) exploration of safety perception. Subsequently, Dubey et al. (2016) pioneered the application of deep learning methods in this field, expanding environmental perception dimensions to safety, lively, boring, wealthy, depressing, and beautiful. With the rapid development of deep learning technology, an increasing number of studies have begun to utilize this technology to analyze environmental elements and their characteristics in street view images (Li et al. 2022; Zhang et al. 2024). In terms of environmental perception, research has primarily focused on people's subjective experience of the environment (Ogawa et al. 2024). Since environmental perception varies with people's social and cultural contexts, Yao et al. (2019) proposed a human-machine adversarial framework and developed a specialized dataset by inviting residents familiar with the local environment to score. Liang et al. (2024) studied the spatial distribution of perceptual attributes of building facades and their impact on overall streetscape evaluation. Zhu et al. (2025) quantified the perceptual characteristics of rural environments across 118 counties in China along five dimensions: wealthy, lively, habitable, tidy, and terroir. In terms of environmental assessment, research emphasizes quantitative analysis based on objective criteria. Xia et al. (2021) developed the Panoramic View Green View Index (PVGVI) using semantic segmentation technology, enabling automated assessment and visualization of urban street greening levels. Additionally, Rui and Cheng (2023) constructed a street environment evaluation system comprising 12 indicators, such as sky visibility index, spatial enclosure index, and road area index, etc., to conduct a quantitative assessment of street landscape quality in the main urban area of Fuzhou City. The integration of street-view imagery and deep learning enables both subjective perception and objective assessment of the environment, providing a new paradigm for urban-rural environmental studies (Wu et al. 2023; Wang et al. 2025).

However, due to the sparse built-up areas and vast geographical spans in rural areas (Li et al. 2020), it is difficult to obtain large-scale and densely distributed rural street view images (Xu et al. 2022). As illustrated in Figure 1, the comparative analysis between the same village's street-view imagery and drone photograph reveals that even a collection of multiple street-view images fails to capture the village's holistic environmental cognition. However, the unique perspective of drones provides more effective visual clues for comprehensive perception and assessment of the rural environment (Kleinschroth et al. 2022). Drone photography is currently widely used in fields such as precision agriculture and tree species monitoring (Buchelt et al. 2024; Epifani et al. 2024). There is also extensive research in traffic monitoring (Kujawski et al. 2021), including pedestrian and vehicle flow analysis, as well as military target recognition (Wang et al. 2024). However, these studies primarily focus on the perception of individual elements, leaving a significant gap in holistic environmental cognition (Zhang et al. 2018). This research gap can be attributed to the inherent complexity of large-scale drone photographs. The abundant environmental features and spatial information embedded in drone photographs pose significant challenges for traditional deep learning models to interpret and process effectively. Notably, the emergence of multimodal large language models (MLLMs) presents new possibilities for achieving holistic perception of large-scale environments.

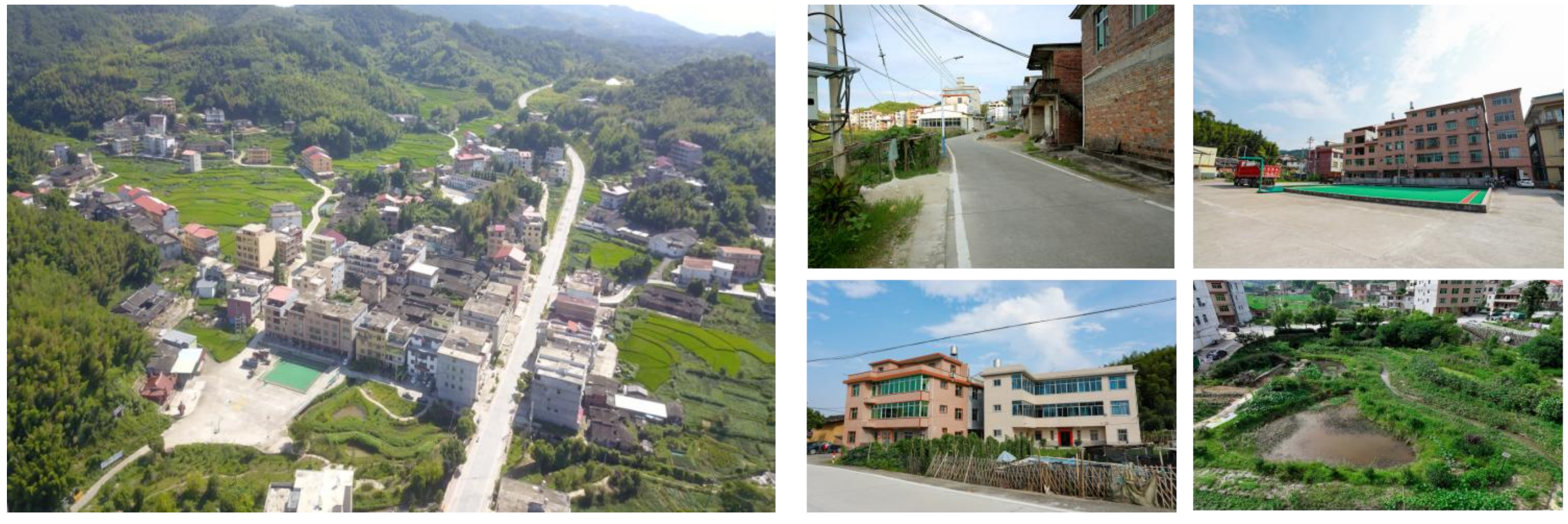

**Fig. 1.** Comparison of ultra-large-scale drone photograph and discrete rural street-view images: Longjin Village, Fujian (Left: drone photograph; Right: corresponding street-view image set)

By deeply aligning the semantic relationships between text and images, MLLMs not only accurately describe image content but also provide insights into the socioeconomic attributes underlying the scene (Wang et al. 2024; Chen et al. 2025). They demonstrate a level of holistic cognition and reasoning capabilities comparable to that of humans (Li et al. 2024). In the domain of urban environmental perception, Zhang et al. (2025) innovatively integrated MLLMs with street-view image analysis, demonstrating their practical value in urban security perception assessment. In complex drone application scenarios, Areerob et al. (2024) leverage extensive expert evaluation data, verified that MLLMs are capable of geological hazard identification and risk assessment based on drone imagery. Comparative studies conducted by Kadiyala et al. (2024) and Samuel et al. (2024) revealed that MLLMs, particularly GPT-4 Vision, have attained near-expert-level analytical capabilities in practical domains, including flood management and water quality monitoring. MLLMs not only enable intelligent analysis of drone-captured imagery but can also be directly deployed on drone platforms (Wu et al. 2025). This integration empowers drones with comprehensive autonomous capabilities, spanning scene perception, spatial reasoning, navigation exploration, task planning, and motion decision-making (Dutta et al. 2024; Ping et al. 2025). This further demonstrates that the cognitive and reasoning capabilities of MLLMs are now sufficient for precise environmental perception and assessment.

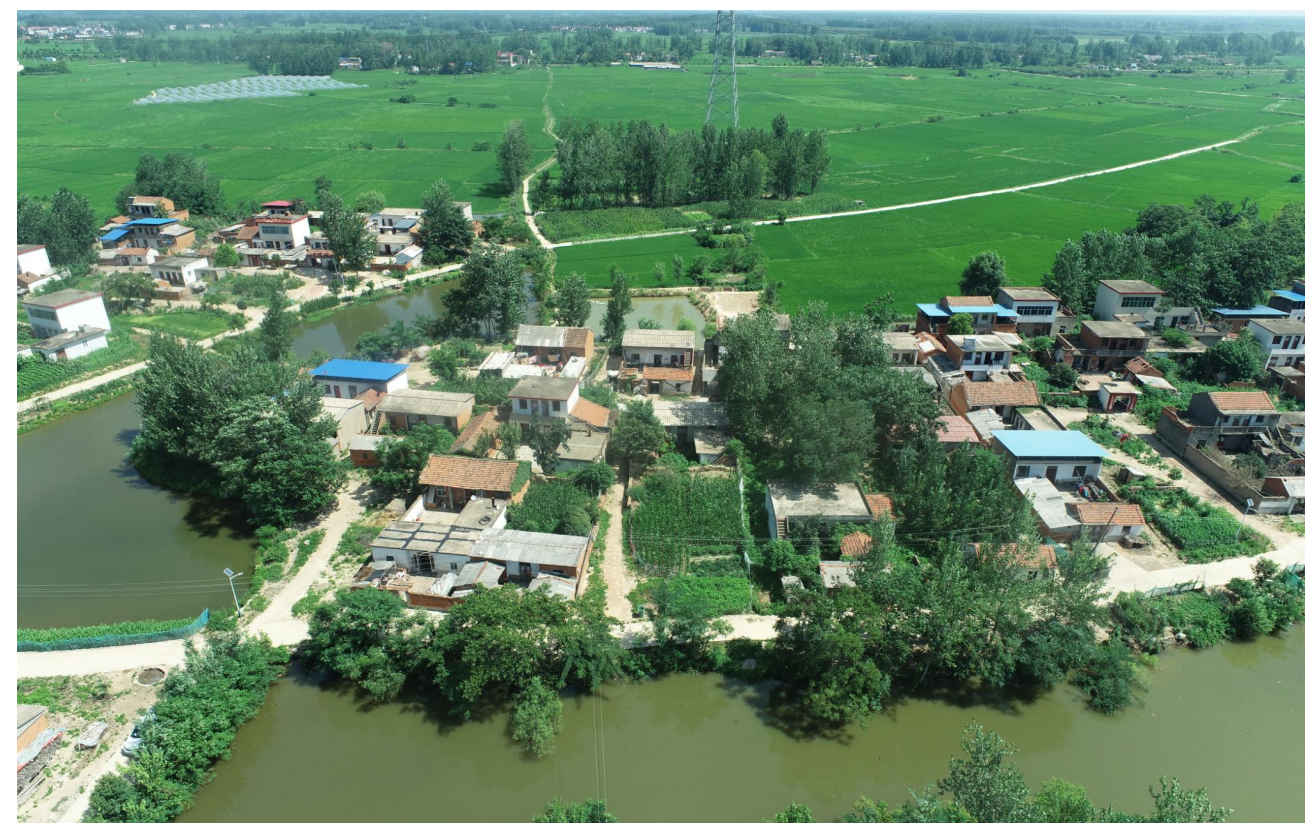

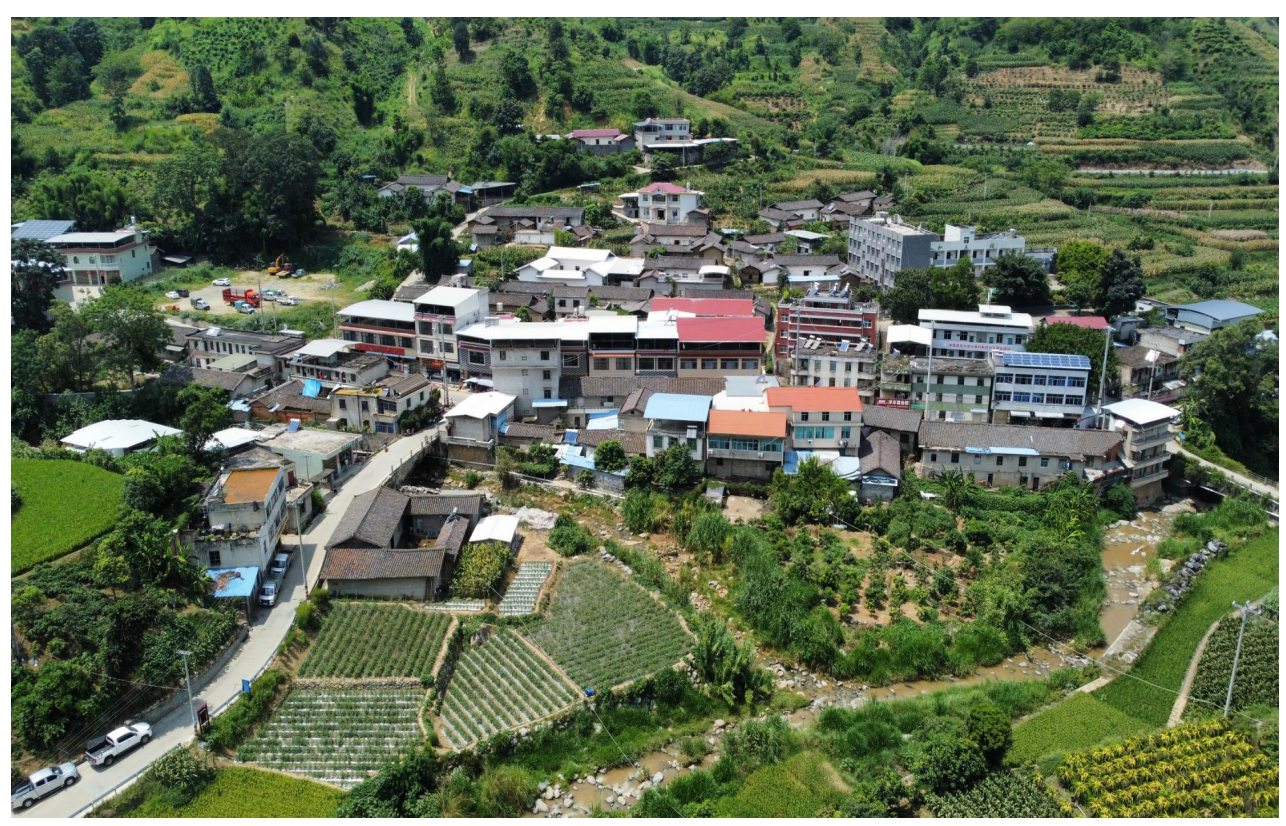

**Fig. 2.** The left side shows a village endowed with a relatively attractive natural landscape, whereas the right side depicts a village featuring more modern living conditions.

While MLLMs demonstrate remarkable general capabilities, they still face significant limitations in specialized environmental perception tasks (Li et al. 2021). As shown in the drone photographs of two villages in Figure 2, most MLLMs excessively rely on visual aesthetics while overlooking critical factors such as housing conditions, road infrastructure, and public facilities that truly determine quality of life (Ma et al. 2020). This leads to significant bias in determining which village offers better livability. Moreover, when analyzing and comparing multiple large-scale drone photographs, excessive information input will cause the MLLMs to produce unstable results, such as confused reasoning and hallucinations.

To address these issues, a visual-language comparative ranking framework (VLCR) is proposed for rural livability assessment. This research fills the gap in using drone photographs for large-scale rural environmental perception and provides a new technical path for rural livability assessment. Specifically, a database of drone photographs covering 1,766 villages in 146 counties nationwide is first constructed, with each image covering a complete settlement unit. To address the limitations of general-domain MLLMs in rural livability assessment tasks, this research proposes the following improvement strategies: First, a chain-of-thought prompting mechanism based on expert knowledge is constructed to facilitate systematic understanding and quantification of quality-of-life and ecological habitability dimensions. Second, an image comparison mechanism with textual description constraints is developed to resolve model instabilities when evaluating livability differences between villages. Third, to address the computational complexity of pairwise comparisons among 1,766 villages, a binary search interpolation ranking algorithm is designed to reduce the required number of comparisons while maintaining assessment accuracy. Based on the above methodological innovations and experimental analysis of nationwide drone photographs, this study successfully characterizes the spatial distribution patterns of rural livability in China. The robustness and reliability of the large-scale rural livability assessment results are further validated through regression analysis against field survey data.

The remainder of this article is structured as follows: Section 2 presents a visual-language comparative ranking framework for quantifying the livability of rural China. Section 3 details the experimental results and provides a comprehensive analysis. Section 4 offers a thorough discussion of the findings. Finally, Section 5 concludes the article.

## 2. Methodology

As an important carrier for recording environmental information, the observation scale of images directly determines the depth and breadth of information expression. Drone photographs, with their unique advantage of combining macro-scale perspectives and micro-level details, provide an ideal data foundation for rural living environment assessment. However, interpreting the complex spatial information in drone photographs requires more advanced analytical tools. In recent years, multimodal large language models (MLLMs) have made significant progress in the field of visual perception (Ren et al. 2023). However, they still have limitations in the deep understanding and reasoning of environmental features. For one thing, the model's interpretation of environmental elements has not yet reached the level of expert cognition; for another, its analytical ability is weakened when faced with high-resolution and large-scale photographs collected by drones. To this end, this study proposes an expert knowledge-guided Vision-Language Contrastive Ranking framework (VLCR), which enhances the environmental perception capabilities of MLLMs to enable systematic assessment of rural livability. Ultimately, this study characterizes the spatial patterns of rural livability across China. The reliability of the national livability prediction results is further validated through field survey data and multiple regression models. The overall workflow of China's rural livability assessment is shown in Figure 3.

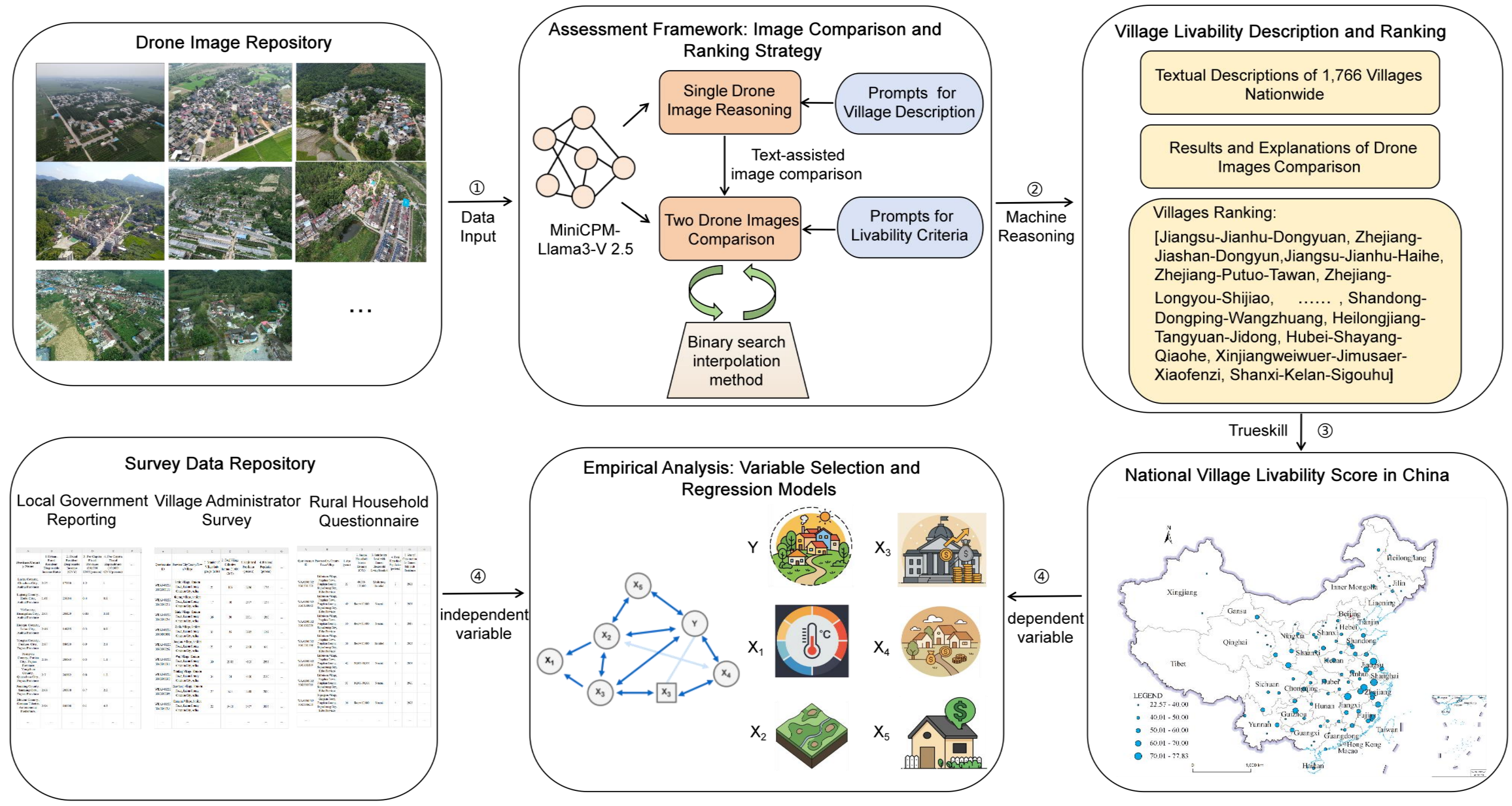

**Fig. 3.** The overall workflow of rural livability assessment in China based on drone photographs and an expert knowledge-guided Vision-Language Contrastive Ranking framework.

### *2.1. Data Repository: Drone photographs and Survey Data*

Based on the National Rural Construction Assessment Project, this study established a comprehensive drone photograph database of rural settlements across China. Data collection adopted an innovative crowdsourcing approach by mobilizing and training village officials to participate in field photography following standardized operational protocols. Considering practicality and cost-effectiveness, consumer-grade drones were utilized, focusing on scene integrity and image clarity rather than professional surveying-grade pixel precision. This grassroots participatory, low-cost data collection model provides a viable pathway for long-term accumulation and sustainable updates of aerial imagery of rural settlements. The database encompasses 1,766 villages across 146 counties, with 111 counties having drone imagery for more than 10 villages. A sufficient number of villages in each county is beneficial for subsequent analysis of differences in impact mechanisms across different regions. To maintain assessment consistency, a single representative panoramic image encompassing the complete morphological characteristics was selected for each rural settlement.

The field survey data in this study are based on a stratified sampling approach, covering 102 sample counties with 10 representative villages selected from each county according to their development levels (high, medium, and low). The data collection consists of three dimensions. At the county level, local governments reported socioeconomic statistical data on per capita fiscal revenue and expenditure, per capita urban and rural residents' savings deposits, per capita GDP, and other county-level development indicators. At the village level, in-depth interviews were conducted with village administrators to collect data on collective income, cultivated land utilization, population structure and mobility, environmental governance, and other village development metrics. At the household level, questionnaire surveys were administered to a minimum of 10 randomly selected households per village to gather data on household income levels, housing modernization conditions, accessibility to amenities, and other relevant indicators. The survey data were primarily utilized to validate the reliability of livability prediction results. Accordingly, this study only selected fiscal expenditure, village collective income, and household annual income as key independent variables. Additionally, to control the impact of environmental factors, this study obtained the annual average temperature and elevation data of all sample counties and used them as control variables.

### 2.2. *Assessment Framework: Drone Photographs Comparison and Villages Ranking Strategy*

This study utilized the MiniCPM-Llama3-V2.5 (Hu et al. 2024) as the experimental benchmark. Differing from other vision fine-tuning models, it possesses three distinct technical advantages. First, it incorporates Scalable Training Strategies designed for small language models, enabling its training outcomes to rival those of large parameter models. Second, the latest RLAIF-V alignment technique has been leveraged in this approach. This innovative methodology has fundamentally altered the previously established paradigm of combining MLLMs with feedback from expert instructors. Instead, a shift has been implemented towards aligning feedback with peers of comparable competence. As a consequence of this transformation, the trustworthiness of the model's responses has been significantly enhanced, while the rate of hallucination has been demonstrably reduced. Additionally, leveraging the cross-language generalization technology of VisCPM, MiniCPM-Llama3-V2.5 has successfully achieved the transfer of large visual models trained in English to Chinese scenarios in a zero-shot manner. With the unique structure and operational mechanism of Vision Transformer, MiniCPM-Llama3-V2.5 is enabled to process images of arbitrary sizes and aspect ratios. This capability facilitates the rapid and efficient interpretation of drone imagery.

Although this model demonstrates powerful image interpretation and inference capabilities, it is characterized by inherent instability and a tendency to score or rate image content based on its own standards. Consequently, the evaluative results of MLLMs for drone photographs cannot be subjected to lateral comparisons. In addition, extensive experiments found that MiniCPM-Llama3-V2.5 frequently confuses the relative positions of two drone photographs, resulting in inconsistent comparison outcomes. To address these issues, absolute scoring of images has been eschewed in this study. Instead, a relative comparison approach has been adopted to obtain more stable and accurate evaluation results. Additionally, a novel drone photographs comparison method and village ranking strategy are designed. The overall process is illustrated in Figure 4.

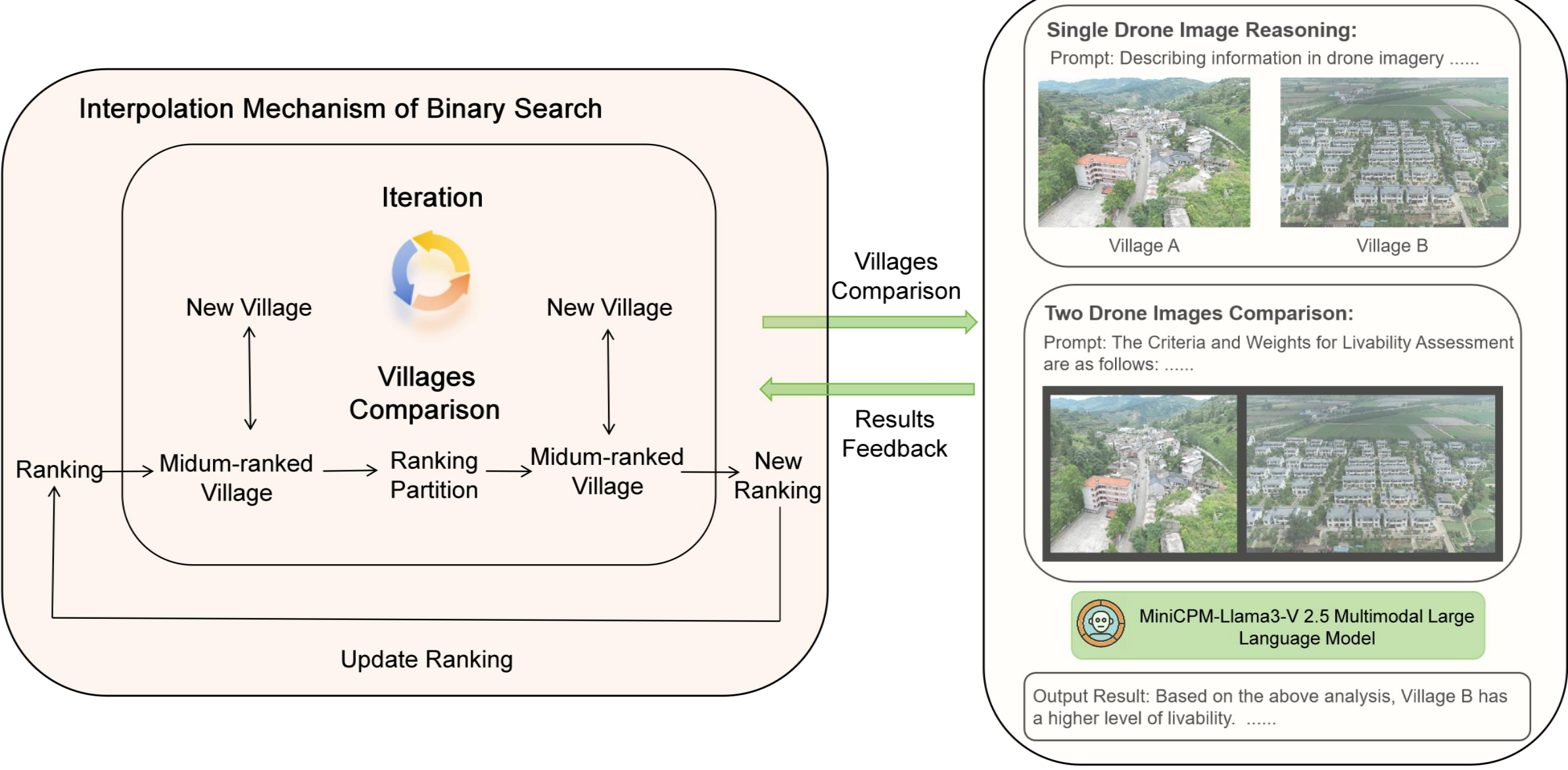

**Fig. 4.** The overall flowchart of drone photographs comparison and village ranking

As illustrated in Figure 4, a text-assisted image comparison methodology has been devised to guide MLLMs in understanding and evaluating village livability. The methodology encompasses several key steps: Firstly, the MiniCPM-Llama3-V-2.5 is employed to generate textual descriptions for each drone photograph, establishing precise correspondence between the photograph and village identifiers. These descriptions serve as characteristic markers for each village, laying the foundation for subsequent comparative assessment. Secondly, this study establishes an expert-knowledge-oriented evaluation framework via chain-of-thought prompting, substantially enhancing the authenticity and accuracy of livability assessments. This framework encompasses six core dimensions: the quality of rural housing construction, the level of infrastructure development, the harmony between buildings and the natural environment, and the amenity of the surrounding landscape, among others. These professional evaluation criteria are transformed into structured prompts with differential weightings, enabling the MLLM to assess village livability across multiple dimensions like an expert. The detailed assessment criteria and their respective weight distributions are presented in Table 1. Finally, based on the concatenated image pairs and their corresponding village identifiers, the model integrates the expert knowledge to generate standardized comparative results of livability in the format of ‘village A > village B’, facilitating subsequent statistical analysis. To ensure the reliability of the assessment results, each village pair undergoes two independent evaluations. If the results are consistent, they are accepted; if inconsistent, a third comparison is conducted, and the final result is determined through majority voting.

Table 1. The Evaluation Criteria and Weightings for Rural Settlement Livability

| Dimensions | Weights |
|---|---|
| The villages are predominantly characterized by row houses of two or more stories. The building heights are observed to be moderate, and the facades are maintained in a clean and uniform manner. A harmonious color scheme is employed throughout. Modern architectural styles and decorative elements are frequently incorporated. | 20% |
| The village development respects natural topography, featuring nature-adaptive building patterns (structures thoughtfully positioned along terrain contours), clear water bodies (transparent streams and ponds with visible riparian vegetation), and large patches of flourishing farmland integrated with surrounding natural greenery. | 20% |
| The village roads are observed to be clean and undamaged, with vibrant coloration. No loose soil is detected on the surfaces. Traffic signs are clearly visible and legible. The main thoroughfares are found to be accessible to vehicular traffic. It is noted that the roads connecting residential areas have been subjected to hardening processes. | 20% |
| The quality of villages is generally observed to be superior where a higher proportion of newly constructed dwellings is present and wall surfaces are found to be undamaged. However, villages predominantly composed of older structures are not necessarily deemed inferior. The assessment of such villages is contingent upon the degree of wear and tear exhibited by the buildings. | 20% |
| The villages feature well-maintained architectural aesthetics, with freshly painted walls and tastefully decorated facades, indicating higher residential livability. | 10% |
| The villages present thoughtfully organized building layouts, characterized by clear structural patterns and orderly orientations, demonstrating a superior living environment. | 10% |

The foundation of village livability assessment is pairwise comparison, but selecting appropriate comparison pairs while minimizing their number remains challenging. Currently, mainstream methods predominantly employ random selection of comparison objects. A complete comparison would require $C_n^2$ comparisons, while most studies typically sample only 20% ~ 30% of all possible pairs to reduce computational costs. However, this approach presents two significant limitations: first, the insufficient number of comparisons may lead to unstable sorting results; second, the computational complexity grows exponentially with the increasing number of villages. Additionally, the integration of new villages into the existing evaluation framework poses challenges in determining appropriate comparison counterparts. To address these challenges, this study proposes a novel method that integrates simultaneous comparison and ranking. The proposed method employs a binary search interpolation algorithm (see Algorithm 1) to optimize comparison frequency while maintaining ranking stability.

**Algorithm 1:** The pseudocode of binary search interpolation

**Input:** $R_0$ : Initial sorting of 20 villages
$R_u$ : Updated rankings
$V_i$ : The i-th village
$V_n$ : All villages to be evaluated
**Output:** $R_n$ : The final village livability ranking
$R_u \leftarrow R_0$
**for** $V_i$ **in** $V_n$ **do**
  $R_u \leftarrow$ BinarySearchInsert( $R_u$ , $V_i$ , $0$, $len(R_u)-1$ )
**end for**
$R_n \leftarrow R_u$
**return** $R_n$
**Function** BinarySearchInsert( $R_u$ , $V_i$ , *left*, *right*):
  $mid \leftarrow \lfloor (left + right) / 2 \rfloor$
  $winner \leftarrow CompareVillages([V_i, R_u[mid]])$
  **if** $winner = V_i$ **then**
    **if** $mid = 0$ **or** $CompareVillages([V_i, R_u[mid-1]]) = R_u[mid-1]$ **then**
      **return** $R_u[:mid] + [V_i] + R_u[mid:]$
    **end if**
    **return** BinarySearchInsert( $R_u$ , $V_i$ , *left*, $mid-1$ )
  **else**
    **if** $mid = len(R_u)-1$ **or** $CompareVillages([V_i, R_u[mid+1]]) = V_i$ **then**
      **return** $R_u[:mid+1] + [V_i] + R_u[mid+1:]$
    **end if**
    **return** BinarySearchInsert( $R_u$ , $V_i$ , $mid+1$, *right*)
  **end if**

With the advantage of binary search interpolation, each comparison only needs to be made with the image at the middle of the existing sorting, requiring at most $\lceil log_2 n \rceil + 1$ comparisons, where ***n*** represents the number of existing sorted images. The formula is $BinarySearchInsert(R_{u-1}, V_i, left, right)$, where $R_u$ represents the updated village ranking, $R_{u-1}$ denote the ranking before the update, and $V_i$ is the village to be ranked. $left$ indicates a higher ranking, and $right$ represents a lower ranking. This algorithm significantly reduces the number of comparisons. To ensure stable results, the order of 20 villages was predetermined $R_0$. Based on this, new village drone photograph $V_i$ are continually compared with the images at the binary positions $mid = \lfloor (left + right)/2 \rfloor$, where $\lfloor \ \rfloor$ denotes the floor symbol. Based on the evaluation results, the algorithm determines whether the village under assessment falls into the upper or lower half of the intermediate position, thereby halving the $left + right$ in each iteration. The process continues iteratively until $left \geq right$, where $left$ and $right$ are dynamically updated variables initialized at 0 and $len(R_{u-1}) - 1$, respectively. The algorithm ultimately determines the optimal ranking position for the village under evaluation, with the return value encompassing two scenarios: $R_u[:mid] + [V_i] + R_u[mid:]$ and $R_u[:mid+1] + [V_i] + R_u[mid+1:]$. The new village is then inserted into the sequence to generate a new sorting $R_u$, and this process is repeated until all villages have been compared. Given that each village can serve as both a comparison object and a compared object, this method also ensures that each village can be compared at least 20 times. Ultimately, 1,766 village descriptions, 16,741 livability comparison result descriptions, and the final village rankings were generated. Compared to conventional ranking methods, our proposed approach demonstrates a threefold improvement in ranking efficiency. As rankings alone cannot adequately illustrate the degree of disparity in village livability, the Trueskill method was employed in this study to convert all comparison results into scores, thereby more intuitively quantifying the differences in village livability levels. Each village was initialized with a baseline score of 25 and a standard deviation of 6, with the final score fluctuation constrained within $\mu \pm 3\sigma$. To highlight regional disparities, scores were standardized and spatially mapped to reveal the distributional patterns of rural livability nationwide.

Moreover, this paper adopts Spearman's Footrule $F(\sigma,\tau)$ as the accuracy evaluation indicator to quantify the difference between two rankings, as delineated in Equation 1. It is employed to compute the sum of absolute differences in the positions of elements across two rankings. This measure is characterized by its simplicity and intuitiveness, and is frequently applied in the comparison of ranking similarities or dissimilarities. In the equation, $F$ denotes the Footrule distance, $\sigma$ and $\tau$ represent the two rankings to be compared, $i$ is utilized as the index of elements, and $\sum$ signifies the summation across all elements.

$$F(\sigma,\tau) = \sum \left| \sigma(i) - \tau(i) \right| \tag{1}$$

### 2.3. Statistical Analysis: Variable Selection and Regression Models

An empirical study was conducted using data collected from questionnaires and local reports. To validate the reliability of national village livability predictions, regression analysis models were employed to examine the relationship between predicted livability scores and various influencing factors. The model evaluated the significance of geographical factors as fundamental determinants, while separately assessing the impacts of state policies, village collective vitality, and individual villagers' behaviors. Firstly, prosperous human settlements have historically developed in areas with favorable climate conditions and terrain features. This suggests that village livability scores can be fundamentally attributed to geographical environmental factors, with environmental variations potentially explaining differences in predicted livability levels. Furthermore, previous studies have widely recognized that villages undergo continuous development, with their livability scores being shaped by multiple stakeholders, including the state, the village collective, and individual villagers. Therefore, a regression model is constructed to examine the relationship between predicted livability scores and these key influencing factors, thus validating the reliability of our prediction framework.

$$F(Liv) = f(Geo) + f(Sta) + f(Col) + f(Vil) \quad (2)$$

$f(Geo)$ is considered the baseline model in this study. The habitability of a region is closely associated with its geographical environment. In this research, temperature (***Tem***) and terrain (***Ter***) are utilized as the control variables to characterize the geographical conditions. Given that both excessively low and high temperatures can impact comfort levels, a quadratic term ($\boldsymbol{Tem^2}$) is incorporated into the regression analysis when examining the influence of temperature. This specification helps capture the inverted U-shaped relationship between temperature and livability, where moderate temperatures typically provide optimal living conditions while extreme temperatures reduce it.

The state power, represented by $f(Sta)$, is directly associated with the livability of regions through governmental investments in infrastructure and public services. In this study, fiscal expenditure (***Fin***) is utilized to reflect this aspect of state force.

The collective force, denoted as $f(Col)$, is representative of the collective strength and reflects the role played by the village collective economic development level and governance capacity in promoting improvements to the village's living environment. In this study, this factor is represented through the utilization of village collective income (***CInc***).

This function $f(Vil)$ represents the impact of villagers' living standards on village livability. The quality of rural housing construction and living convenience indirectly reflect the villages' livability level.

The villagers' income (***VInc***) is primarily selected to reflect this aspect.

$$Y = \beta_0 + \beta_1 Tem^2 + \beta_2 Tem + \beta_3 Ter + \beta_4 Fin + \beta_5 CInc + \beta_6 VInc + e \quad (3)$$

The potential multicollinearity among our independent variables (temperature, terrain, governmental fiscal investments, village collective economic revenues, and individual household income) was carefully examined. We conducted a comprehensive multicollinearity test through Variance Inflation Factor (VIF) analysis. The results show that all VIF values are well below the conventional threshold of 10, indicating no significant multicollinearity exists among these variables in explaining rural livability.

## 3. Experiments Results

### *3.1. Implementation Details*

Compared to the traditional deep learning models, Multimodal Large Models (MLLMs) demonstrated significant advantages in zero-shot image perception and reasoning. Additionally, drone photographs, as one of the important windows reflecting rural landscapes, have been widely proven to effectively present the overall appearance of villages. Therefore, this study utilized the state-of-the-art MiniCPM-Llama3-V2.5 as a pipeline to conduct livability assessments on 1,766 villages in 146 counties across 28 provinces in China based on drone aerial photographs. The study did not fine-tune the model with image instructions but rather carefully designed a chain-of-thought prompting to let the MLLM evaluate rural livability based on expert standards and cognition.

To mitigate the risk of errors, fabrications, or inaccuracies in content generation by MLLMs, the Temperature parameter of the model was set to 0.2 in this study. Moreover, the number of comparisons for each pair of images was increased to 3. These measures were implemented to enhance the reliability and consistency of the generated results. To ensure the acquisition of complete output, the max-tokens parameter was configured to 1024. During the process of image contrast analysis, a horizontal juxtaposition method was employed, wherein two images to be compared were combined side by side into a single larger image. A novel text-assisted strategy was specifically designed to enable the large model to accurately distinguish between the left and right images, thereby improving the stability and accuracy of the comparative analysis. During the image concatenation process, the following key parameters were established: a 500-pixel gap between the two images and an 80-pixel border width. Additionally, to ensure the applicability of the final composite image, the maximum resolution of the images is limited to 3060 pixels in width and 1440 pixels in height. Moreover, all experiments were conducted in parallel on a high-performance server equipped with two GeForce RTX 4090 GPUs, providing a total of 48GB of available GPU memory.

### *3.2. Implementation results and comparisons*

To validate the effectiveness of the proposed strategy, this study selected drone images from 20 villages in various provinces across the country. These images were then separately subjected to comparative experiments on the currently strongest open-source and commercially available MLLMs, including Llama3-llava-next-8b (Liu et al. 2024), Claude-3-5-sonnet (Anthropic 2024), Gemini 1.5 Pro (Team et al. 2024), gpt-4-vision-preview (Achiam et al. 2023), MiniCPM-Llama3-V2.5 (Hu et al. 2024), and our framework VLCR. The resulting rankings were compared against expert rankings to verify the effectiveness of our designed framework and strategy. The experimental results are presented in Table 2.

Table 2. Comparative Analysis of MLLMs Rankings and Expert Rankings on Village Livability: Ground Truth, Llama3-llava-next-8b, MiniCPM-Llama3-V2.5, VLCR (our), Claude-3-5-sonnet, Gemini 1.5 Pro, and gpt-4-vision-preview

| Sorting | Ground Truth (GT) | Llama3-llava-next-8b (w/o strategy) | Llama3-llava-next-8b (w/ strategy) | MiniCPM-Llama3-V2.5 (w/o strategy) | VLCR (our) | Claude-3-5-sonnet | Gemini 1.5 Pro | gpt-4-vision-preview |
|---|---|---|---|---|---|---|---|---|
| 1 | Jiangsu-Jianhe | Xinjiang-Quanshuidi | Hunan-Qianjian | Sichuan-Tounian | Zhejiang-Xinan | Fujian-Longjin | Zhejiang-Xinan | Zhejiang-Xinan |
| 2 | Zhejiang-Xinan | Fujian-Longjin | Hubei-Qiaotoubian | Henan-Hunan | Fujian-Longjin | Zhejiang-Xinan | Fujian-Longjin | Jiangsu-Jianhe |
| 3 | Anhui-Kandong | Hubei-Qiaotoubian | Zhejiang-Xinan | Hubei-Qiaotoubian | Jiangsu-Jianhe | Henan-Hunan | Jiangsu-Jianhe | Fujian-Longjin |

| | | | | | | | |
|---|---|---|---|---|---|---|---|
| 4 | Fujian-Longjin | Hunan-Qianjian | Guangdong-Zhankeng | Jiangxi-Wangjia | Shaanxi-Dishuiya | Shandong-Liuyao | Hubei-Qiaotoubian | Hubei-Qiaotoubian |
| 5 | Jiangxi-Wangjia | Guangxi-Caijia | Liaoning-Hongfeng | Shandong-Liuyao | Hunan-Qianjian | Sichuan-Tounian | Shandong-Liuyao | Sichuan-Tounian |
| 6 | Sichuan-Tounian | Yunnan-Xiaoguan | Sichuan-Tounian | Zhejiang-Xinan | Liaoning-Hongfeng | Jiangsu-Jianhe | Guangdong-Zhankeng | Guangdong-Zhankeng |
| 7 | Hubei-Qiaotoubian | Shandong-Liuyao | Shaanxi-Dishuiya | Gansu-Wenchi | Anhui-Kandong | Liaoning-Hongfeng | Hunan-Qianjian | Anhui-Kandong |
| 8 | Shaanxi-Dishuiya | Hainan-Sanduo | Jiangsu-Jianhe | Jiangsu-Jianhe | Jiangxi-Wangjia | Guangdong Zhankeng | Anhui-Kandong | Henan-Hunan |
| 9 | Henan-Hunan | Guangdong-Zhankeng | Jilin-Xinfeng | Jilin-Xinfeng | Sichuan-Tounian | Jilin-Xinfeng | Liaoning-Hongfeng | Hunan-Qianjian |
| 10 | Hunan-Qianjian | Qinghai-Shangkewa | Jiangxi-Wangjia | Guangdong-Zhankeng | Shandong-Liuyao | Hubei-Qiaotoubian | Henan-Hunan | Jilin-Xinfeng |
| 11 | Guangdong-Zhankeng | Gansu-Wenchi | Gansu-Wenchi | Liaoning-Hongfeng | Henan-Hunan | Shaanxi-Dishuiya | Sichuan-Tounian | Liaoning-Hongfeng |
| 12 | Liaoning-Hongfeng | Jiangxi-Wangjia | Qinghai-Shangkewa | Yunnan-Xiaoguan | Hubei-Qiaotoubian | Hainan-Sanduo | Xinjiang-Quanshuidi | Shandong-Liuyao |
| 13 | Gansu-Wenchi | Jiangsu-Jianhe | Hainan-Sanduo | Fujian-Longjin | Guangdong-Zhankeng | Gansu-Wenchi | Gansu-Wenchi | Hainan-Sanduo |
| 14 | Shandong-Liuyao | Sichuan-Tounian | Fujian-Longjin | Shaanxi-Dishuiya | Gansu-Wenchi | Jiangxi-Wangjia | Jilin-Xinfeng | Qinghai-Shangkewa |
| 15 | Hainan-Sanduo | Henan-Hunan | Anhui-Kandong | Anhui-Kandong | Hainan-Sanduo | Hunan-Qianjian | Hainan-Sanduo | Xinjiang-Quanshuidi |
| 16 | Guangxi-Caijia | Liaoning-Hongfeng | Henan-Hunan | Qinghai-Shangkewa | Jilin-Xinfeng | Yunnan-Xiaoguan | Jiangxi-Wangjia | Gansu-Wenchi |
| 17 | Jilin-Xinfeng | Zhejiang-Xinan | Shandong-Liuyao | Hainan-Sanduo | Xinjiang-Quanshuidi | Anhui-Kandong | Shaanxi-Dishuiya | Jiangxi-Wangjia |
| 18 | Qinghai-Shangkewa | Anhui-Kandong | Yunnan-Xiaoguan | Xinjiang-Quanshuidi | Qinghai-Shangkewa | Xinjiang-Quanshuidi | Qinghai-Shangkewa | Yunnan-Xiaoguan |
| 19 | Xinjiang-Quanshuidi | Jilin-Xinfeng | Guangxi-Caijia | Guangxi-Caijia | Yunnan-Xiaoguan | Guangxi-Caijia | Yunnan-Xiaoguan | Shaanxi-Dishuiya |
| 20 | Yunnan-Xiaoguan | Shaanxi-Dishuiya | Xinjiang-Quanshuidi | Hunan-Qianjian | Guangxi-Caijia | Qinghai-Shangkewa | Guangxi-Caijia | Guangxi-Caijia |
| **F(σ,τ)** | | 0.19 | 0.51 | 0.47 | **0.74** | 0.56 | 0.63 | 0.65 |

The Ground Truth in the table represented a comprehensive result obtained through evaluations by a multidisciplinary team of experts. This assessment team consisted of scholars from relevant fields such as human geography, urban and rural planning, and landscape design, who collaboratively formed the final rankings through their assessments. Considering the diversity of expertise within the evaluation team and the rigor of the assessment process, this study regarded these ranking results as highly credible and valuable for reference. For the open-source Llama3-llava-next-8b and MiniCPM-Llama3-V2.5 models, a pairwise image comparison method and text-assisted strategy were employed for evaluation. In contrast, the commercial MLLMs such as Claude-3-5-sonnet, Gemini 1.5 Pro, and gpt-4-vision-preview had specific requirements for the size and quantity of input images: the image size generally should not exceed 5MB, and the number of images processed in a single batch should not exceed 20. It was found through extensive experiments that 10 images were the most suitable quantity for these models to process. Based on this finding, 5 sets of images, each containing 10 images, were carefully selected to generate the ranking results of 20 images.

Based on the evaluation results from various methods, it can be observed that the output generated by our VLCR framework exhibited the highest similarity to the Ground Truth, with a similarity of 0.74, but this value may fluctuate. This was followed by gpt-4-vision-preview and Gemini 1.5 Pro, with similarities of 0.65 and 0.63 respectively. Upon implementation of the image comparison strategy proposed in this study, the MiniCPM-Llama3-V2.5 model not only demonstrated a 0.27 increase in similarity, but also surpassed commercial MLLMs in terms of evaluation accuracy. These findings effectively illustrated the efficacy and potential of the proposed strategy. It is worth noting that before applying our proposed strategy, the Llama3-llava-next-8b model achieved a similarity of only 0.19 with the Ground Truth, and exhibited unstable performance. However, following the introduction of the image comparison strategy, its similarity to the Ground Truth significantly improved to 0.51. The substantial improvement in accuracy and stability has demonstrated the universality and effectiveness of the proposed strategy.

The comparison results show that most MLLMs tend to prioritize the aesthetic harmony of the natural environment, such as the overall layout of mountains, waters, forests, fields, lakes, and grasslands, when assessing village livability. As evidenced by Table 2, mainstream commercial MLLMs generally perceive Hunan Village in Henan Province as having high livability. However, drone imagery reveals that the actual housing conditions and road construction in this village are not ideal, as illustrated on the left side of Figure 2. In light of these findings, the village comparison strategy proposed in this study has adjusted the evaluation criteria for village livability, emphasizing infrastructure, housing conditions, and living environment as key assessment dimensions. Tounian Village in Sichuan Province has well-developed infrastructure, many large public buildings, and newly constructed rural houses, as shown on the right side of Figure 2. Taking these characteristics into consideration, our method has classified this village as a settlement with a relatively high level of livability. To facilitate a visual comparison of livability differences across villages, Figure 5 presents drone imagery of villages with different livability scores.

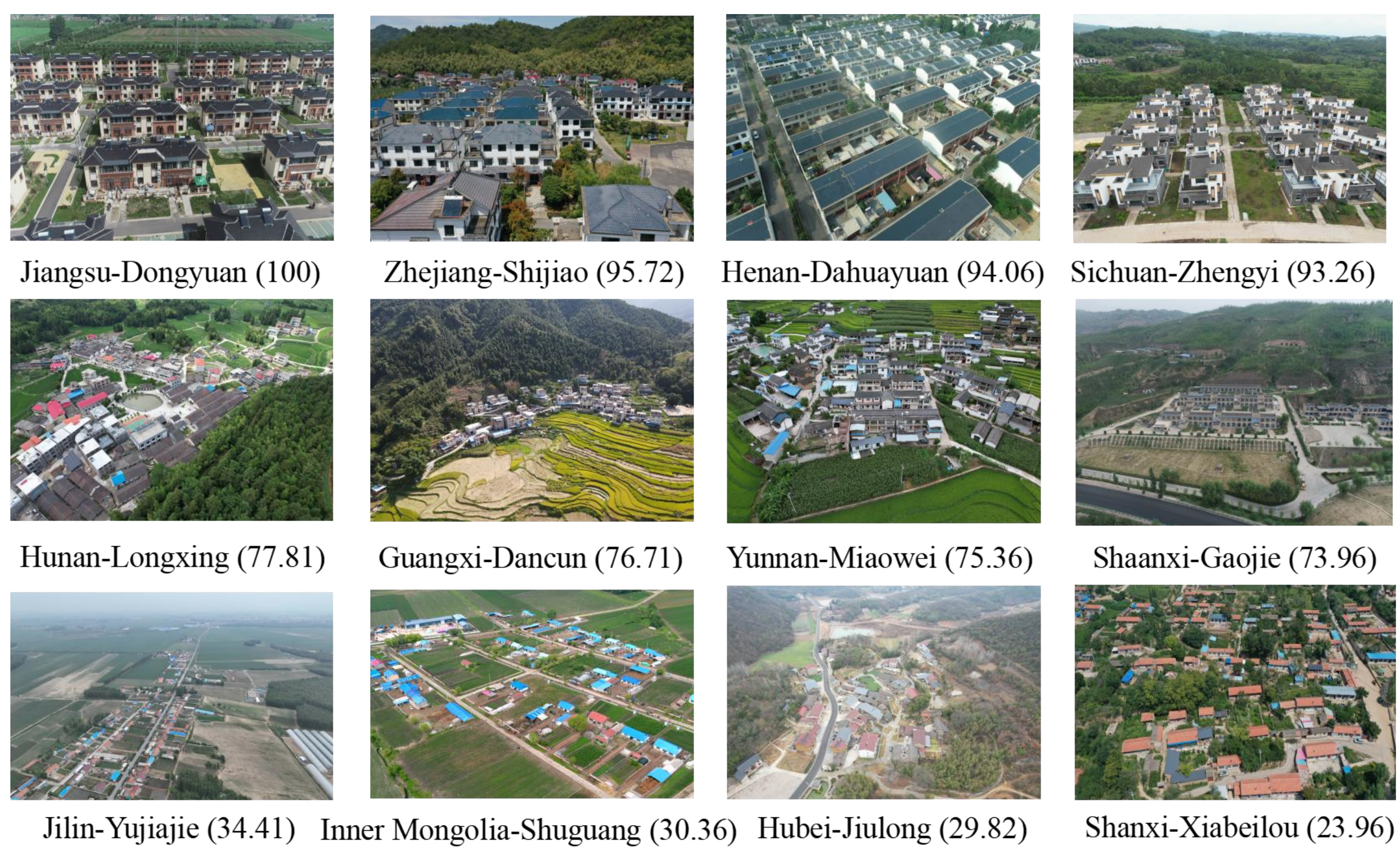

**Fig. 5.** Drone imagery comparison of villages with different livability scores

### *3.3. Spatial Differentiation Patterns of Rural Livability*

Village livability was systematically evaluated nationwide through the VLCR framework. The regional disparity pattern of villages' livability scores across the country is shown in Figure 6. Statistical analysis revealed that the average livability score for villages nationwide was 46.28, indicating substantial room for improvement in overall livability. The eastern coastal regions demonstrated higher livability, with Zhejiang ranking first at 64.92, followed by Jiangsu and Fujian at 58.98 and 53.35, respectively. These regions are characterized by developed economies, well-established infrastructure, and favorable climatic conditions, rendering them more suitable for habitation and development. Central regions exhibited moderate livability, with Anhui, Hunan, Hubei, and Henan scoring between 45 and 50. While these areas boast diverse topography and distinct seasonal climates, their livability scores are lower than those of the eastern coastal regions, possibly due to relatively lower levels of economic development and government investment.

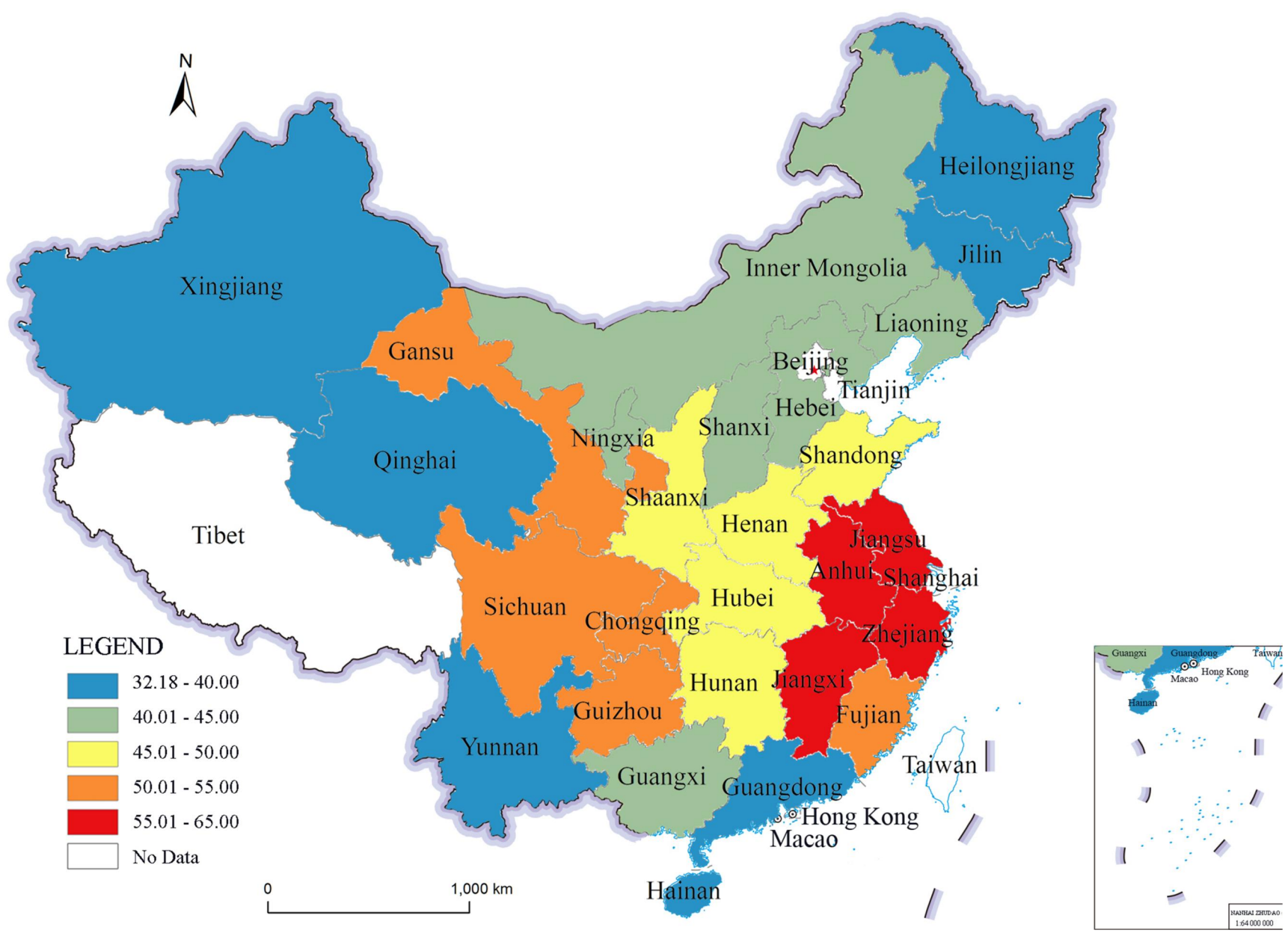

**Fig. 6.** The regional disparity pattern of villages' livability levels at the provincial scale

Some areas in the western region have relatively high livability, including Gansu, Guizhou, Sichuan, and Chongqing. Although these regions have complex terrain and relatively lagging economic development, substantial funds have been invested in construction through the Western Development Strategy and Rural Revitalization Strategy. As a result, the living environment has significantly improved. The northeastern and border regions of China, including Heilongjiang, Jilin, and Qinghai, demonstrate relatively low livability scores. It can be attributed to the harsh climate conditions and predominant plateau-mountainous terrain, which significantly constrain the development of village livability. In addition, this regional difference in China also reflects the significant impact of economic development on village livability. The eastern coastal regions have experienced rapid economic growth due to their first-mover advantage in China's Reform and Opening-up policy. This economic prosperity has generated substantial fiscal capacity to support rural infrastructure development. Additionally, the higher income levels of rural residents in these areas have led to greater emphasis and investment in improving living environments, further enhancing village livability.

An interesting finding from our analysis shows that, despite being a major labor-exporting province with relatively low economic development, Jiangxi demonstrates remarkably high rural livability levels. Field investigations reveal that Jiangxi residents tend to invest their earnings in rural housing improvements, contributing to highly livable rural landscapes. In contrast, Guangdong Province, despite being China's leading economic powerhouse, shows relatively low rural livability scores. This can be attributed to the great differences in economic development among regions within Guangdong Province. Our sample data primarily focused on counties in western and northern Guangdong, which are generally less developed regions, thus resulting in lower overall livability assessments.

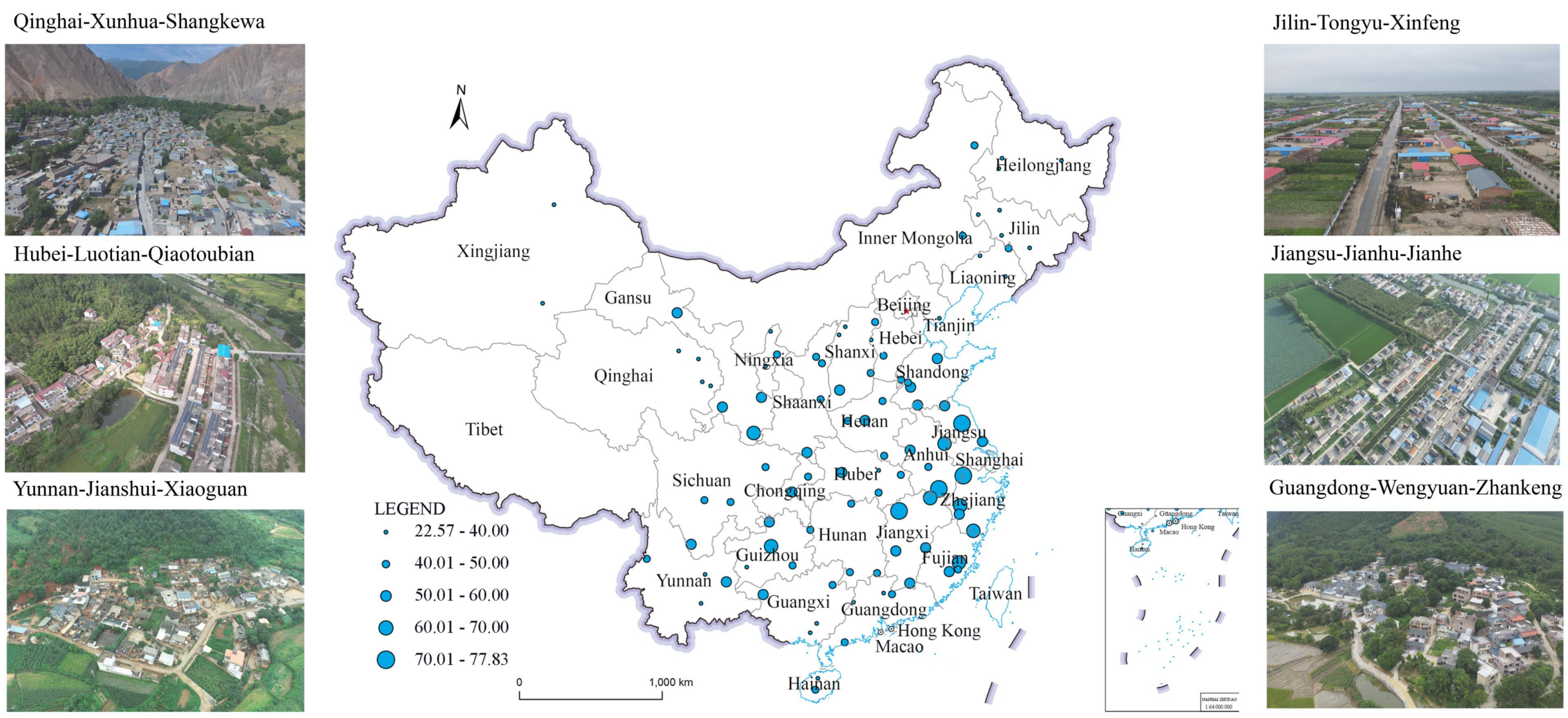

**Fig.7.** The distribution pattern of village livability scores at the county scale

From the county-scale results, significant regional disparities are similarly reflected, as shown in Figure 7. The livability scores of counties in the middle and lower reaches of the Yangtze River are observed to be relatively high. For instance, Jianhu County in Jiangsu Province ranks among the top with a score of 77.83, while Yi County in Anhui Province and Anji County in Zhejiang Province score 71.71 and 70.19, respectively. The middle and lower reaches of the Yangtze River are characterized by plains and river networks, where the superior natural geographical environment is demonstrated to contribute to the higher residential attractiveness of these regions.

Counties in the three northeastern provinces are generally found to have lower scores. For example, Tangyuan County in Heilongjiang Province is scored at merely 22.57, Benxi County in Liaoning Province at 32.93, and Huinan County in Jilin Province at 33.33. These low scores can be attributed to the cold climate and prolonged winters associated with high-latitude regions, factors by which people's living experiences and quality of life are adversely affected. It is noteworthy that the northeastern region has been predominantly characterized by population outflow in recent years.

In the central and western regions, a marked differentiation is observed. Some areas located in mountainous terrain, such as Guanling County in Guizhou Province, are scored as low as 38.62, while others situated in valley plains, like Kaiyang County in Guizhou, achieve a relatively high score of 64.93. The majority of counties in these regions are located in mountainous or plateau areas, with complex topographical environments and significant development challenges. As a result, the villages in these areas primarily rely on state investments to improve their living environments and quality of life. This rural development pattern has further led to unbalanced development within the region.

# 4. Discussion

## 4.1. Methodological Contributions

Previous studies on rural livability primarily rely on questionnaire surveys and rural street view perception. However, the former faces limitations in large-scale applications, while the latter lacks credibility due to insufficient image sampling density. To achieve an objective and quantitative measurement of rural livability, this study innovatively designs a measurement system that utilizes landscape as the perspective, drone photographs as the data source, and multimodal large models (MLLMs) as the analytical tool. The drone photographs offer a meso-scale representation of the villages, containing comprehensive information about the entire village. It provides an ideal depiction of rural landscapes. Moreover, MLLMs possess powerful capabilities in image perception and reasoning, allowing them to understand landscape information in a way that resembles human cognition. Utilizing expert-knowledge-oriented chain-of-thought prompts, these models can systematically analyze and evaluate villages across multiple dimensions, effectively identifying their strengths and limitations. Furthermore, the design of binary search interpolation algorithms significantly enhances the efficiency of drone photograph comparison and sorting processes. This evaluation system not only achieves a transition from qualitative to quantitative research but also fills a gap in assessing rural livability through drone photographs.

## 4.2. Statistical Analysis

To validate the reliability of the national livability prediction results, regression analyses are performed to examine the relationships between livability results and three social forces: state force, collective force, and villagers' force. The results not only aligned well with theoretical expectations but also quantified the specific impact of each social force on rural livability, further supporting the validity of our assessment methodology. Specifically, this study collects questionnaire data from 102 sample counties, selecting fiscal expenditure, village collective income, and household income as the proxy variables for state force, collective force, and villagers' force. Considering the impact of the natural environment on village livability, this study selected temperature and terrain as control variables. Fiscal expenditure, village collective income, and household income are designated as explanatory variables, with village livability level as the dependent variable. Ultimately, a multiple regression model for village livability level is constructed.

Table 3. Results of Multiple Regression Analysis between Livability Scores and Key Variables: Temperature, Topography, Fiscal Expenditure, Village Collective Income, and Household Income

| | Dependent variable: | | | | |
|---|---|---|---|---|---|
| | (1) | (2) | (3) | (4) | (5) |
| ***Tem²*** | -0.106*** | -0.125*** | -0.105*** | -0.100*** | -0.118*** |
| ***Tem*** | 3.082*** | 3.569*** | 3.131*** | 2.686*** | 3.314*** |
| ***Ter*** | 0.049** | 0.028 | 0.075*** | 0.076*** | 0.060** |
| ***Fin*** | | 4.908*** | | | 3.920*** |
| ***CInc*** | | | 0.104*** | | 0.066* |
| ***Inc*** | | | | 8.313*** | 3.801 |
| Constant | 23.891*** | 15.125*** | 14.355*** | 11.515** | 5.279 |
| Observations | 94 | 92 | 94 | 94 | 92 |
| $R^2$ | 0.248 | 0.369 | 0.326 | 0.348 | 0.439 |
| Adjusted $R^2$ | 0.223 | 0.340 | 0.295 | 0.319 | 0.400 |
| Residual Std. Error | 9.536 (df = 90) | 8.858 (df = 87) | 9.080 (df = 89) | 8.928 (df = 89) | 8.450 (df = 85) |
| F Statistic | 9.886*** (df = 3; 90) | 12.726*** (df = 4; 87) | 10.747*** (df = 4; 89) | 11.879*** (df = 4; 89) | 11.092*** (df = 6; 85) |
| Note: | *p<0.1; **p<0.05; ***p<0.01 | | | | |

Models (2) - (4) are designed to examine the impacts of government investment, collective vitality, and villagers' behavior on village livability, respectively, while controlling for geographical environmental influences. Model (2) demonstrates that government fiscal investment has a significant positive effect on village livability, with the explanatory power reaching 36.9%. It is observed that higher government investment correlates with improved village livability. Analysis of county-level data reveals that the average per capita fiscal investment required is 16,900 yuan/(person·year). Eastern regions attain 19,300 yuan, while northeastern regions only reach 13,100 yuan. This pattern generally aligns with the regional disparities in village livability observed nationwide.

Model (3) investigates the impact of village collective economic income on promoting village livability, while controlling for geographical environmental factors. China's rural areas have long operated under a collective land ownership system, with many village affairs being collectively deliberated and funded. Therefore, the vitality of the village collective directly affects the level of village development. The results from Model (3) indicate that counties with higher levels of village collective economic strength exhibit greater village livability. The explanatory power of Model (3) reaches 32.6%, slightly lower than that of Model (2).

Model (4) examines the effect of household income on village livability, yielding a significantly positive result. However, in Model (5), which comprehensively analyzes the roles of the state, collective, and villagers while controlling for geographical environmental influences, villagers' income is found to be insignificant. This suggests that villagers are more concerned with matters directly related to themselves. Their personal income is often inclined to be invested in housing construction, while relatively less is contributed to village development. In contrast, investments made by the state and village collectives are found to have a direct impact on village livability levels.

### *4.3. Limitations and future work*

This study also has several limitations. Firstly, this framework was tested with three repetitions at a temperature of 0.2, yielding Spearman's Footrule Distances of 0.74, 0.62, and 0.64, respectively. The experimental results exhibit a certain degree of instability, which necessitates further in-depth investigation to explore the underlying causes. Secondly, regarding the livability evaluation criteria, our assessment framework primarily addresses physical dimensions such as housing modernization levels, road infrastructure quality, and complementary public service facilities. However, we acknowledge that this expert-based evaluation framework does not encompass crucial cultural elements, including historical heritage, traditional landscapes, and local characteristics that contribute to community identity and place attachment. Future research should expand the evaluation dimensions to develop a more inclusive and comprehensive assessment of village livability. Finally, this study primarily focuses on objective indicators, without fully incorporating residents' subjective satisfaction and personal preferences. Future research will incorporate questionnaire data to develop a multidimensional evaluation framework, thereby more accurately reflecting the actual livability of villages.

## 5. Conclusion

This study develops an innovative framework for evaluating village livability in China based on drone photographs and multimodal large language models. To address the potential image-text misalignment issue within the model, a novel text-assisted contrastive strategy is introduced. Extensive experiments demonstrated that the proposed strategy, which integrates carefully designed chain-of-thought prompts, significantly enhances both the stability and accuracy of image comparison results. Furthermore, a ranking mechanism based on binary search interpolation is designed in this study, enabling parallel processing of image comparison and ranking, thereby markedly improving ranking efficiency.

Research indicates that the livability of rural areas in China exhibits significant regional differences and hierarchical characteristics. The eastern coastal regions reveal the highest level of livability, followed by the western regions, forming a spatial pattern that radiates outward from these centers with decreasing livability. The northeast and northwest regions have the lowest livability levels. In addition, the results of multiple regression analysis show that the predicted scores of national village livability are significantly correlated with the natural environment, fiscal expenditure, village collective income, and household income. Notably, fiscal expenditure demonstrates the strongest impact among all factors, which aligns with our field survey observations. These findings further validate the robustness and generalizability of our methodology.

**Declaration of competing interest**

The authors declare that they have no known competing financial interests or personal relationships that could have appeared to influence the work reported in this paper.

**Data availability statement**

The data and code in the paper will be released at https://huggingface.co/dwh/RLA-MLMs.

**Acknowledgement**

The authors would like to thank the editor, the associate editor, and the anonymous reviewers for their helpful comments and advice. This work was supported by the National Natural Science Foundation of China (Grant No. 42371206).